\documentclass[aps,prb,reprint,twocolumn,showpacs,superscriptaddress]{revtex4-1}
\usepackage{amsmath, amssymb, amsthm}
\usepackage{graphicx}

\usepackage[usenames,dvipsnames]{color}			
\definecolor{purple}{rgb}{0.5,0,0.5}

\def\hall{\sigma^{12}_{xy}}

\begin{document}

\title{Monte Carlo study of phase transitions out of Symmetry-Enriched Topological phases of bosons in two dimensions}
\author{Jong Yeon Lee}
\affiliation{Department of Physics, California Institute of Technology, Pasadena, California 91125, USA}
\affiliation{Institute for Quantum Information and Matter, Caltech, Pasadena, California 91125, USA}
\author{Scott Geraedts}
\affiliation{Department of Physics, California Institute of Technology, Pasadena, California 91125, USA}
\affiliation{Institute for Quantum Information and Matter, Caltech, Pasadena, California 91125, USA}
\author{Olexei I. Motrunich}
\affiliation{Department of Physics, California Institute of Technology, Pasadena, California 91125, USA}
\affiliation{Institute for Quantum Information and Matter, Caltech, Pasadena, California 91125, USA}

\date{\today}

\begin{abstract}
We study a statistical mechanics model of two species of bosons with mutual statistics $\theta=2\pi/n$ in (2+1) dimensions. This model realizes a fractionalized topological phase of bosons, which is a fractionalized version of the boson integer quantum Hall effect. The model can be studied with sign-free Monte Carlo simulations. We study the phase transitions between the fractionalized topological phase and a trivial insulator, and between different topological phases. We find that these transitions are continuous, and we measure their critical exponents. 
\end{abstract}

\maketitle

\section{Introduction}

From the time the notion of topological order was introduced,\cite{Wen1990} topological quantum phases have attracted much attention from physicists. One defining property of such phases is that they admit quasiparticles with fractionalized charge and statistics (anyons). Examples include quasiparticles in the fractional quantum Hall effect,\cite{Stern2008} spinon and vison excitations in $Z_2$ spin liquids,\cite{Read1989,Kitaev2003,SenthilFisher_Z2,Wen91} and excitations in a variety of fractionalized systems such as string-net liquids.\cite{Levin2005,Nayak2008_rmp,MotrunichZ3,LevinStern2009} It is also natural to ask about possible new phases that such particles can have, as a way to access proximate phases and phase transitions involving topological quantum states.\cite{Zhang1989,FisherLee1989,Tupitsyn2010,Barkeshli2010,Kou2009,Wen2000,Burnell2011,Gils2009}

One practical way to study a condensed matter system is to do importance sampling of its imaginary-time path integral via the Monte-Carlo method.  However, anyonic statistics can give rise to complex values in the path integral, causing an infamous ``sign problem'' that prevents direct Monte Carlo studies.  Therefore, in a previous work,\cite{Loopy} we developed a model with two species of loops and \emph{mutual statistics} that can be reformulated in a sign-free form. Each species of loop represents the worldline of a particle, which is a boson with respect to particles of the same species.  The two species of particles are \emph{mutual anyons}, having mutual statistics $\theta$, which gives rise to interesting phenomena.
Previous work\cite{FQHE} also showed that closely related models realize the ``integer quantum Hall effect for bosons'' phase proposed by Lu and Vishwanath.\cite{LuVishwanath} This is a symmetry-protected topological phase protected by a $U(1)$ charge conservation symmetry, and has a Hall conductivity quantized to an even integer. Our models with mutual statistics realize fractionalized versions of such a phase---so-called symmetry-enriched topological (SET) phases. These phases have a Hall conductivity which is quantized to values equal to two times a rational number, and they have quasiparticles carrying fractional charge and mutual statistics. We will call these fractionalized phases ``fractional quantum Hall insulators'' (FQHI). Unlike the conventional fractional quantum Hall effect in strong magnetic field such as Laughlin states, these SET phases are not chiral and require a $U(1)$ symmetry.

Such concrete models can be a powerful tool for studying topological phases. In this paper we will study phase transitions from FQHI phases to a trivial insulator, and between two different FQHI phases. Our Monte Carlo simulations will allow us to determine the order of these transitions and to extract critical exponents.

\begin{figure}[b]
	\vspace{-0.2in}
	\includegraphics[width = 1.15\columnwidth]{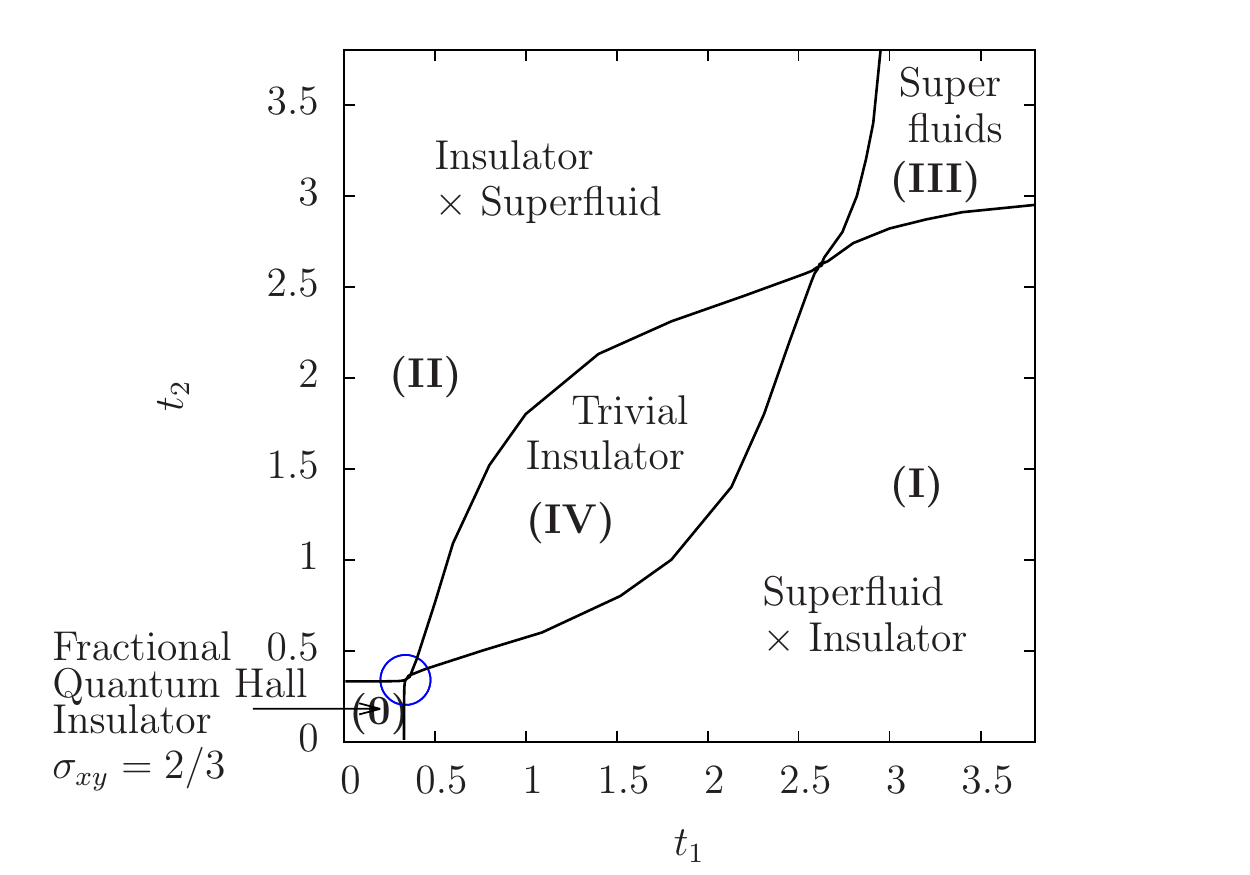}
	\caption{\label{fig:phase} The phase diagram for the model with $\theta=2\pi/3$. In the left bottom corner is the fractionalized phase. In terms of physical bosons, this phase has fractional Hall conductivity and gapped quasiparticles which are mutual anyons. The blue circle centered at $(0.34, 0.34)$ represents the critical region of our interest. In the other phases the two species of physical bosons are essentially uncorrelated: each species can be in either a trivial insulating or superfluid phase, leading to four different topologically-trivial phases.}
\end{figure}

The model used in this work is defined by the following action:
\begin{equation}
S \left[ \vec{J}_1, \vec{J}_2 \right]  = \sum_r \frac{\vec{J}_1(r)^2}{2t_1} + \sum_R \frac{\vec{J}_2(R)^2}{2t_2}
+ i \theta \sum_r \vec{J}_1(r) \cdot \vec{p}_2(r) ~. \label{eq:model}
\end{equation}
The index $r$ refers to sites on a cubic lattice (the \emph{direct} lattice) and $R$ refers to sites on another, interpenetrating cubic lattice (the \emph{dual} lattice) whose sites lie at the centers of the cubes of the direct lattice.  $J_{1\mu}(r)$ is an integer-valued current on a link $r, r+\hat{\mu}$ of the direct lattice and $J_{2\mu}(R)$ is another integer-valued current on a link $R, R+\hat{\mu}$ of the dual lattice.  We use schematic vector notation so that $\vec{J}_1$ and $\vec{J}_2$ represent these integer-valued currents. The currents form closed loops, i.e., $\vec{\nabla} \cdot \vec{J}_1 = \vec{\nabla} \cdot \vec{J}_2 = 0$.  In the partition sum, a given current configuration obtains a phase factor $e^{i\theta}$ or $e^{-i\theta}$ for each cross linking of the two loop systems, dependent on the relative orientation of the current loops. This ``mutual statistics'' is realized in the last term of Eq.~(\ref{eq:model}) by including a ``gauge field'' $\vec{p}_2$, defined on the direct lattice, whose flux encodes the $\vec{J}_2$ currents via $\vec{\nabla} \times \vec{p}_2 = \vec{J}_2$, where lattice curl is defined as
\begin{equation}
(\vec{\nabla} \times \vec{p}_2)_\mu(R) = \sum_{\nu,\lambda} \epsilon^{\mu\nu\lambda}[p_\lambda(r + \hat\nu) - p_\lambda(r)] ~.
\end{equation}
Note that in three space-time dimensions, taking the curl of an object on the links of the direct lattice gives an object on the links of the dual lattice. We also require that the currents have zero total winding in our system with periodic boundary conditions; this guarantees that the above model is precisely mathematically defined.\cite{Loopy, short_range3, FQHE} The parameters $t_1$ and $t_2$ give the strength of the on-site interactions felt by the currents. 

In Ref.~\onlinecite{FQHE}, we found a specific two-dimensional (2D) quantum Hamiltonian with short-range interactions that has a phase with gapped quasiparticles described by the above action.  In terms of such a physical 2D Hamiltonian, this phase is a fractionalized phase with $\sigma_{xy} = 2/n$ for $\theta = 2\pi/n$.  Note that in terms of the 2D quantum Hamiltonian, $\vec{J}_1$ and $\vec{J}_2$ currents represent the world-lines of quasiparticle excitations of the fractionalized phase, not the world-lines of the elementary bosons. These quasiparticles are gapped in the fractionalized phase, and condensing them leads to a trivial insulator. In this paper, we focus on this quasiparticle condensation transition.  
In the Appendix, we will show how to identify the elementary bosons (which we will often call physical bosons), starting from the above quasiparticle action.  Whenever we quote $\sigma_{xy}$ to describe phases occuring in our model, we always mean response properties in terms of such physical bosons and physical Hamiltonian.

Figure~\ref{fig:phase} shows the phase diagram for the action in Eq.~(\ref{eq:model}) with $\theta=2\pi/3$, obtained using Monte Carlo simulations.\cite{short_range3}  There are five distinct phases in the phase diagram, which is symmetric with respect to the line $t_1 = t_2$ due to the symmetric nature of the action under $\vec{J}_1 \leftrightarrow \vec{J}_2$ and $t_1 \leftrightarrow t_2$.  As mentioned earlier, we can rewrite Eq.~(\ref{eq:model}) in terms of the elementary bosons of the system, and it is in terms of these bosons that we describe the nature of the phases.  The phase in the center is a trivial insulating phase, while the phase in the left bottom corner is the fractionalized phase with $\sigma_{xy} = 2\frac{1}{3}$.  The phase in the upper right corner is a superfluid phase, where the elementary bosons are proliferated. Phases off the diagonal in the picture are phases in which one species of the physical bosons is in a superfluid phase, while the other is in a trivial insulating phase. Except for the region marked by the blue circle, the phase transition boundaries were studied previously\cite{short_range3} and found to be second order (continuous) phase transitions of the 3D (classical) XY type, corresponding to the condensation of a single species of particles (which are one of the elementary bosons when going from the trivial to superfluid phase or one of the quasiparticles when going from the fractional to superfluid phase).  However, the nature of the critical region inside the blue circle was left unclear and therefore remains to be examined more thoroughly. In this work, we will focus on this phase transition, which tentatively is a transition between a fractional quantum Hall insulator and a trivial insulator.

The outline of the paper is as follows. In Sec.~\ref{sec:MC}, we define the measurements used to study phase transitions in our models, and how they are exploited to extract information about criticality and critical exponents.  Section~\ref{sec:results} contains the results of the Monte Carlo study.  From these results, we deduce that the region of criticality in the blue circle of Fig.~\ref{fig:phase} is a second-order multi-critical point and extract critical exponents. We repeat the same study for several cases where $\theta$ has the form $2\pi/n$ with integer $n$. In Sec.~\ref{sec:general}, we consider phase diagrams for more complicated values of $\theta$ using results developed in the Appendix, and show that different microscopic models give the same local phase diagrams, which can be exploited to study the universality of topological phase transitions. Section~\ref{sec:concl} summarizes our result and discusses further generalizations.

\section{Monte Carlo Method and Measurements}
\label{sec:MC}

The action in Eq.~(\ref{eq:model}) is complex-valued due to the mutual statistics term. Since we want to compute expectation values with respect to this action via the Monte Carlo method, these complex values seem to lead to a sign problem which would cause this method to fail. However, in Ref.~\onlinecite{Loopy} we showed how to reformulate this action by summing over one species of current to obtain a real-valued action. We will not describe the method here, but interested readers can find all relevant information in Ref.~\onlinecite{short_range3}. The resulting sign-free action is expressed in terms of the gauge field $\vec{p}_2$ and the variables $\phi_1$, which represent the boson phases conjugate to the $\vec{J}_1$ variables:
\begin{eqnarray}\label{eq:model2}
&&S\left[ \phi_1(r), \gamma_\mu, \vec{p}_2(r) \right] = \sum_R \frac{[(\vec{\nabla} \times \vec{p}_2)(R)]^2}{2t_2} \\
&&~~ + \sum_{r,\mu} V_{\text{Villain}}[\phi_1(r+\hat{\mu}) - \phi_1(r) -\theta p_{2\mu}(r) - \gamma_\mu \delta_{r_\mu ,0} ], \nonumber
\end{eqnarray}
where $V_{\text{Villain}}$ is a $2\pi$-periodic ``Villain potential", which is obtained by summing over the $\vec{J}_1$ variables. In order to enforce zero total winding in our system, we introduce auxiliary variables $\gamma_\mu \in (-\pi,\pi)$ (operating like fluctuating boundary conditions on the $\phi_1$) which we integrate over to provide the desired constraints. Correlations in terms of the original variables $\vec{J}_1$, $\vec{J}_2$ can be extracted in terms of $\phi_1, \vec{p}_2$.\cite{short_range3} In previous works, by performing Monte Carlo simulations on this reformulation, we determined the phase diagram for $\theta = \pi$ and $\theta = 2\pi/3$. In this work we will perform a similar study, but focusing on the critical region circled in Fig.~\ref{fig:phase}.

In order to study the model in Monte Carlo, we will monitor several thermodynamic variables. First, we monitor internal energy per site $\epsilon \equiv S/\text{Vol}$, where $S$ is the action in the sign-free reformulation that is being simulated and $\text{Vol} = L^3$ is the total number of sites in the system. Internal energy is used to compute ``heat capacity,'' defined as 
\begin{equation}
C \equiv (\langle \epsilon^2 \rangle - \langle \epsilon \rangle^2 ) \times \text{Vol} ~.
\label{heatcap}
\end{equation}
Measuring the heat capacity is useful because we can detect phase transitions by studying its singularities.

To study the behavior of the current variables more directly, we first compute their Fourier transforms:
\begin{equation}
J_{a\mu} (k) \equiv \sum_r \frac{1}{\sqrt{\text{Vol}}} J_{a\mu}(r) e^{-i k \cdot r} ~,
\label{fourier}
\end{equation}
where $k$ is a wave vector and $a = 1, 2$ labels the current species. We then monitor current-current correlations given by
\begin{equation}
C^{ab}_{\mu \nu} = \langle J_{a\mu}(-k_{\rm min}) J_{b\nu}(k_{\rm min}) \rangle ~,
\label{Cabmn}
\end{equation}
where $k_{\rm min} $ is the smallest wave vector in a direction other than that of the currents being measured. For simplicity in this section we will set $k_{\rm min}$ in the $z$-direction, $k_{\rm min} \equiv (0, 0, 2\pi/L)$, and therefore $\mu, \nu \in {x, y}$, though when we show numerical data we have averaged over all directions. We will always measure these correlations at $k_{\rm min}$, and so from now on we omit the $k$ labels when writing these quantities. These current-current correlations can be used to identify the different phases of the system, as their dependence on system size $L$ is different in different phases.
For example, $C^{22}_{xx}$ is the ``superfluid stiffness'' of the $J_2$ bosons and therefore has a non-zero value (independent of the system size) in phases (II) and (III), while it is proportional to $1/L^2$ in the other phases, and $C^{11}_{xx}$ behaves similarly.
Another way to understand this is that our system has $U(1)\times U(1)$ symmetry [one $U(1)$ from each species of conserved bosons], and the stiffnesses $C^{11}_{xx}$ and $C^{22}_{xx}$ detect the breaking of these symmetries.

There is no symmetry breaking in both the trivial insulator and fractional quantum Hall insulator, and so we need other tools to identify these phases. In particular, we can use the cross-species transverse correlation $C^{12}_{xy}$. This quantity can be related to the Hall conductivity of the system,\cite{Gen2Loops,FQHE} and therefore it can be used to identify the fractionalized Hall phases. Due to the symmetry of our model under spatial rotations, $C^{12}_{xx} = C^{11}_{xy} = 0$, and $C^{11}_{xx} = C^{11}_{yy} = C^{11}_{zz}$.  Furthermore, along the symmetric line $t_1 = t_2$, we have $C^{11}_{xx} = C^{22}_{xx}$, etc. Therefore from now on we will omit subscripts and write $C^{22}\equiv C^{22}_{xx}$ and $C^{12}\equiv C^{12}_{xy}$.

One way to determine the critical exponent $\nu$ is to study peaks in the heat capacity. Let $t_{\rm crit}$ be the location of the critical point and $\alpha$ the critical exponent for the heat capacity. Then for a model with parameter $t$, $C \propto |t-t_{\rm crit}|^{-\alpha}$ near a phase transition. Since the correlation length $\xi \propto |t - t_{\rm crit}|^{-\nu}$ near the phase transition, hyperscaling relation in space-time dimension $d$ gives $\alpha = 2 - \nu d$.  (Note that our models are space-time isotropic, so have dynamical exponent $z = 1$.)  Then conventional finite-size scaling analysis gives $C_{\rm peak} \propto L^{(2/\nu - d)}$, where $C_{\rm peak}$ is the height of the peak. Therefore we can in principle use heat capacity to extract the critical exponent $\nu$. However, if $\nu d > 2$, which means $\alpha < 0$, we will have only a cusp singularity and the above scheme using heat capacity does not work.

We will see in the next section that this is the case in our system, and we therefore need other methods to determine critical exponents.  To this end, we compute the derivative of the current-current correlation function $C^{22}(k_{\text{min}}) \cdot L$. 
As will be justified in the next section, $C^{22}$ should have the following scaling form:
\begin{equation} \label{scaling}
C^{22}(k_{\text{min}}) \cdot L = \tilde{f} [ L / \xi ]
= f[L^{1/\nu} (t - t_{\rm crit})] ~,
\end{equation}
where $f(x)$ is some single-variable function of $x$ such that $f(x \rightarrow 0) \rightarrow {\rm const} \neq 0$ and $f(x \rightarrow \infty) \rightarrow 0$ (on either side of the transition).

\begin{figure}
	\includegraphics[width=\columnwidth, angle = 0]{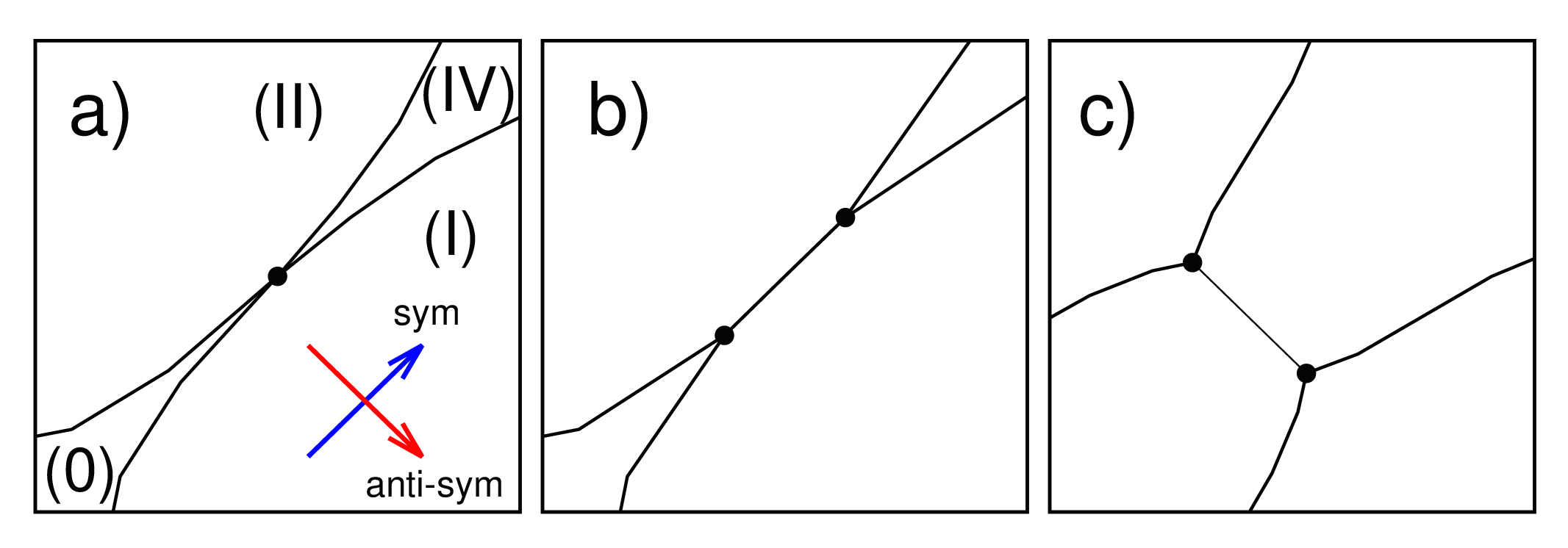}
	\caption{\label{fig:scenarios} Different scenarios for how the phases can meet on the $t_1 = t_2$ line near the transition from the fractionalized phase to the trivial insulator, which is marked by blue circle in Fig.~\ref{fig:phase}. The blue arrow represents the symmetric direction and the red arrow represents the anti-symmetric direction. As argued in the main text, our evidence points to a) being the correct scenario and that the transition is second-order.
}
\end{figure}

The model is symmetric under the exchange of $t_1 \leftrightarrow t_2$ and $\vec{J}_1 \leftrightarrow \vec{J}_2$, so the phase diagram must be symmetric with respect to the line $t_1 = t_2$. We define the symmetric parameter $t_s = (t_1 + t_2)/2$ and anti-symmetric parameter $t_a = (t_1 - t_2)/2$.  The corresponding directions are indicated  in Fig.~\ref{fig:scenarios} with blue and red arrows.

In our model, we expect two exponents $\nu_s$ and $\nu_a$ describing the divergence of the correlation length as we approach the critical point along the symmetric and anitsymmetric directions respectively.
The above scaling behavior of $C^{22} (k_{\text{min}}) \cdot L$ generalized to the case with deviations in both of these directions is:
\begin{equation} \label{scaling}
C^{22}(k_{\rm min}) \cdot L = f(L^{1/\nu_s} \delta t_s, L^{1/\nu_a} \delta t_a) ~.
\end{equation}
The derivative of the above quantity with respect to $t_s$ or $t_a$ evaluated at the critical point will be proportional to $L^{1/\nu_s}$ or $L^{1/\nu_a}$, respectively.

We can measure such symmetric and antisymmetric derivatives of the current-current correlation as follows:
\begin{eqnarray}
\frac{\partial C^{ab}_{\mu\nu}}{\partial t_{s/a}} &=& \frac{\partial C^{ab}_{\mu\nu}}{\partial t_1} \pm \frac{\partial C^{ab}_{\mu\nu}}{\partial t_2} ~, \nonumber \\
\frac{\partial C^{ab}_{\mu\nu}}{\partial t_i} &=& \frac{1}{2t_i^2} \bigg[ \langle J_{a\mu}(-k_{\rm min}) J_{b\nu}(k_{\rm min}) J_i^2  \rangle \nonumber \\
&&~~~~~~ - \langle J_{a\mu}(-k_{\rm min}) J_{b\nu}(k_{\rm min}) \rangle \langle J_i^2 \rangle \bigg] ~.
 \label{eq:deriv}
\end{eqnarray}

We determined the critical exponents in the symmetric and antisymmetric directions, $\nu_s$ and $\nu_a$, by measuring the above quantities near the critical point.

\vspace{-0.1in}


\section{Results}
\label{sec:results}
\subsection{Nature of the Critical Region}

Before identifying the order of the transition indicated in Fig.~\ref{fig:phase}, we investigate the nature of the critical region. In particular, we want to know how the four phases meet. Three possible scenarios\cite{short_range3} are shown in Fig.~\ref{fig:scenarios}. In the figure, (0) stands for the fractional quantum Hall insulator, and (IV) stands for the trivial insulator phase. To study the nature of the critical region, we performed thorough simulations with different system sizes near the critical region inside the blue circle in Fig.~\ref{fig:phase}. In this section we primarily show data for $\theta=2\pi/3$, although we have similar results for several other $\theta$ values.

\begin{figure}[t]
	\includegraphics[width = 1.0\columnwidth, angle = 0]{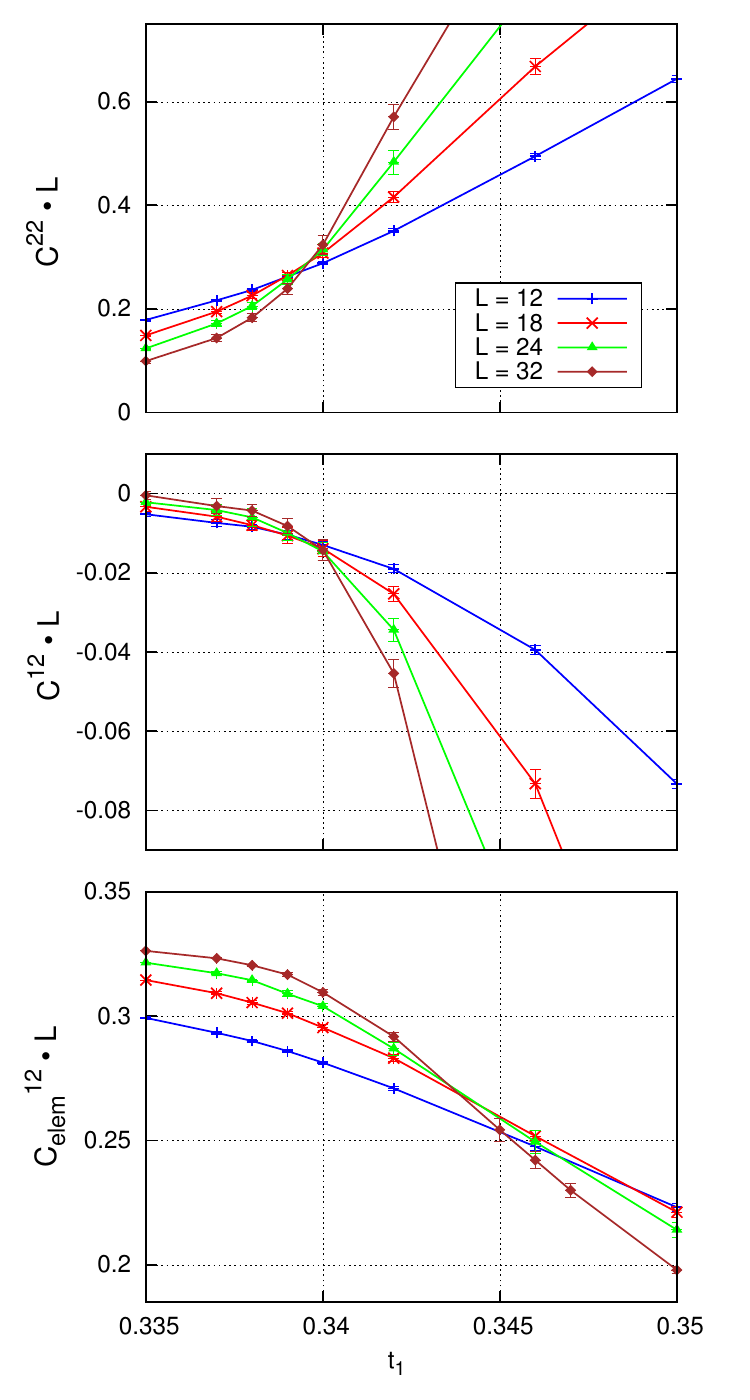}
	\vspace{-0.2in}
	\caption{\label{fig:3} Behavior of the current-current correlations $C^{22} \cdot L$ and $C^{12} \cdot L$ in the range [0.335, 0.350] near the critical point for the transition from the phase (0) to phase (IV), taken along the symmetrical line $t_2 = t_1$.  Error bars are obtained by running 20 simulations with different random seeds.  The crossing points for data with different system sizes allow us to bound the critical region.
Top panel: Current-current correlation for $J_{2x}$ and $J_{2x}$, cf.~Eq.~(\ref{Cabmn}). 
Middle panel: Cross-species transverse current correlation for $J_{1x}$ and $J_{2y}$. 
Bottom panel: Cross-species transverse correlation for currents of the elementary (physical) bosons; see text for details.
}
\end{figure}

We first investigate scenario (a) vs (b), i.e., we try to determine whether the system goes through a critical point or a critical line segment as $t_s$ is increased. 
In the top panel of Fig.~\ref{fig:3} we show how $C^{22}\cdot L$ behaves in the critical region. One clearly sees that at the transition this quantity is changing. Though one might try to determine the nature of the transition from such data, we will soon see that the behavior of this quantity near the critical point is somewhat unusual. Therefore we will instead focus on the inter-species current-current correlation $C^{12}\cdot L$. 
We note that in the fractionalized phase $C^{12}\cdot L\propto 1/L^2$, while we have found (see also Ref.~\onlinecite{short_range3}) that in the trivial insulator phase $C^{12}\cdot L = -3$ in the large system limit.  Though this may seem counterintuitive, remember that the $C^{12}$ is a correlator of quasiparticles of the fractionalized phase and not elementary bosons.  Since $C^{12}\cdot L$ has fixed thermodynamic limit values when we are deep inside each phase, the finite-size $C^{12}\cdot L$ should interpolate between these values near the phase transition. 

The middle panel of Fig.~\ref{fig:3} shows $C^{12}\cdot L$ for different system sizes $L=12,18,24,32$. The lines cross at $t_s\approx 0.340$, with the crossing point moving to larger values as system size is increased. If the transition happens at a single point, such as in scenarios (a) or (c), this crossing approaches the location of that point in the thermodynamic limit. On the other hand, if scenario (b) is correct, we expect that this crossing marks the lower end of the critical segment, even if we do not know the behavior of $C^{12}\cdot L$ along the critical segment.

So far we have discussed correlation functions of the quasiparticle currents $J$, but it is also possible to measure correlation functions of the elementary boson currents, $C^{12}_{\rm elem}$. A sketch of how to do this is provided in the Appendix, with a more general treatment in Ref.~\onlinecite{Gen2Loops}. Note that in the thermodynamic limit the Hall conductivity is given by $2C^{12}_{\rm elem}\cdot L$. Therefore $C^{12}_{\rm elem}\cdot L$ takes the value $1/3$ in the fractionalized phase and zero in the trivial insulator. The lower panel of Fig.~\ref{fig:3} shows a plot of this quantity. Once again there is a crossing between data at different system sizes. This crossing occurs at $t_s\approx 0.344$ for the largest system sizes, and it moves to smaller values as the system size is increased. This value approaches the location of the critical point in scenarios (a) or (c), or the upper end of the critical line segment if (b) is correct. 

\begin{figure}[h]
	\includegraphics[height = 1\columnwidth, angle = 270]{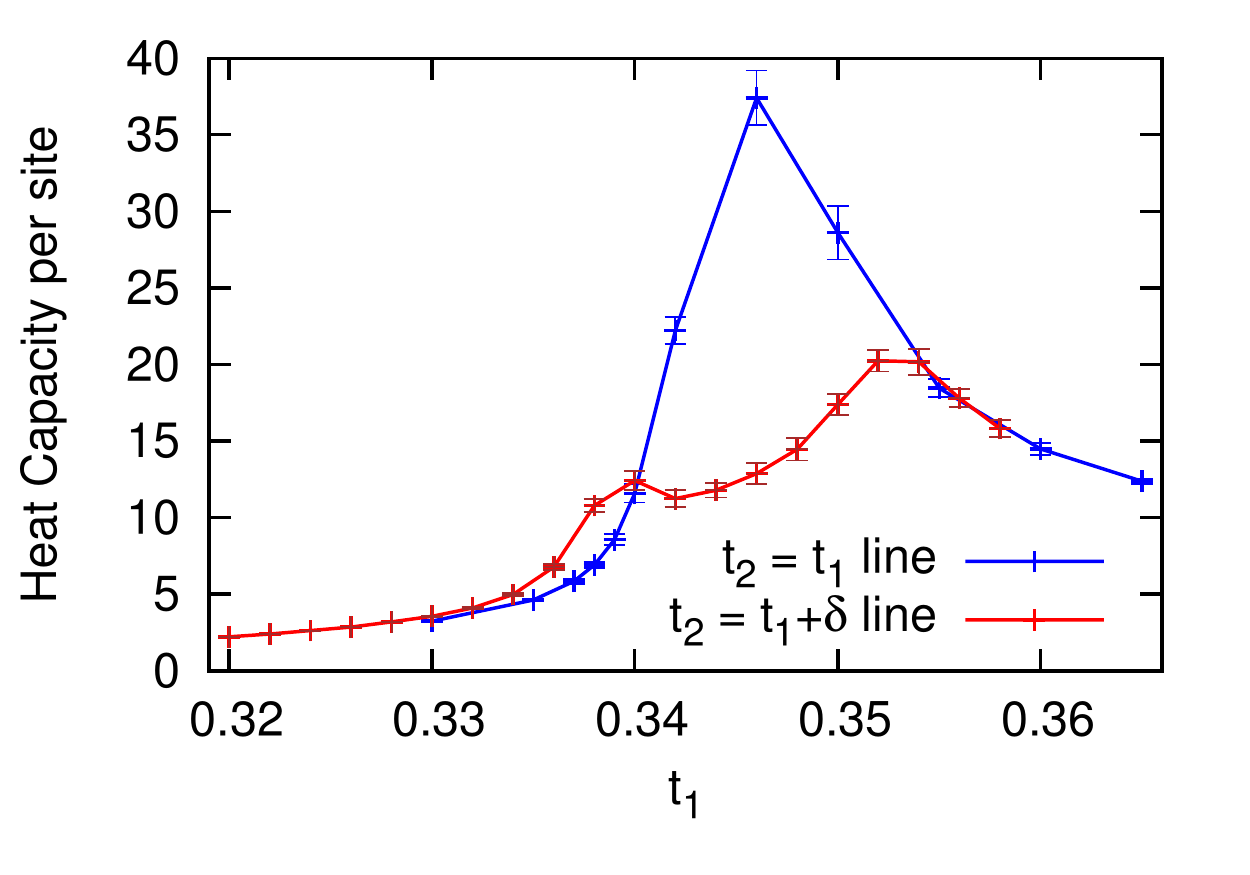}
	\caption{\label{fig:4} Behavior of heat capacity per site, Eq.~(\ref{heatcap}), along the symmetric line ($t_1 = t_2$) and parallel line ($t_1 = t_2 + \delta$, with $\delta = 0.002$) for the model with $\theta = 2\pi/3$ and system size $L = 32$. The heat capacity is single-peaked for the symmetric line and double-peaked for the parallel line. This implies that there are two phase transitions along the parallel line, excluding scenario (c) in Fig.~\ref{fig:scenarios}}.
\end{figure}

If scenario (b) is true, our critical segment will have length $< 0.004$. Considering that all the parameters $t_1$, $t_2$, and $\theta$ have values of order unity, such a very small critical segment length under the scenario (b) is hard to explain. Therefore, we believe that scenario (b) is invalid. A more likely explanation for the difference between the crossing points of $C^{12}\cdot L$ and $C^{12}_{\rm elem}\cdot L$ is that the crossing points will continue to move as system size is increased and they will coincide in the thermodynamic limit.

Now we have to choose between scenarios (a) and (c). If scenario (c) is valid, then along a nearby line parallel to the symmetric line we will come across a phase transition only once, so the heat capacity will be single-peaked along the line. To test this assumption, we sweep along the line $t_2 = t_1 + \delta$, where $\delta = 0.002$. Figure~\ref{fig:4} shows that along this parallel line, the heat capacity actually has two peaks, implying that there are two phase transitions. Therefore, if scenario (c) is correct, then the segment of criticality perpendicular to the symmetric line would have size $< 0.004$. Again, this is a value much smaller than any of the parameters in our model, so we conclude that this scenario is also unlikely.

\begin{figure}[h]
	\includegraphics[height = 0.95\columnwidth, angle =270]{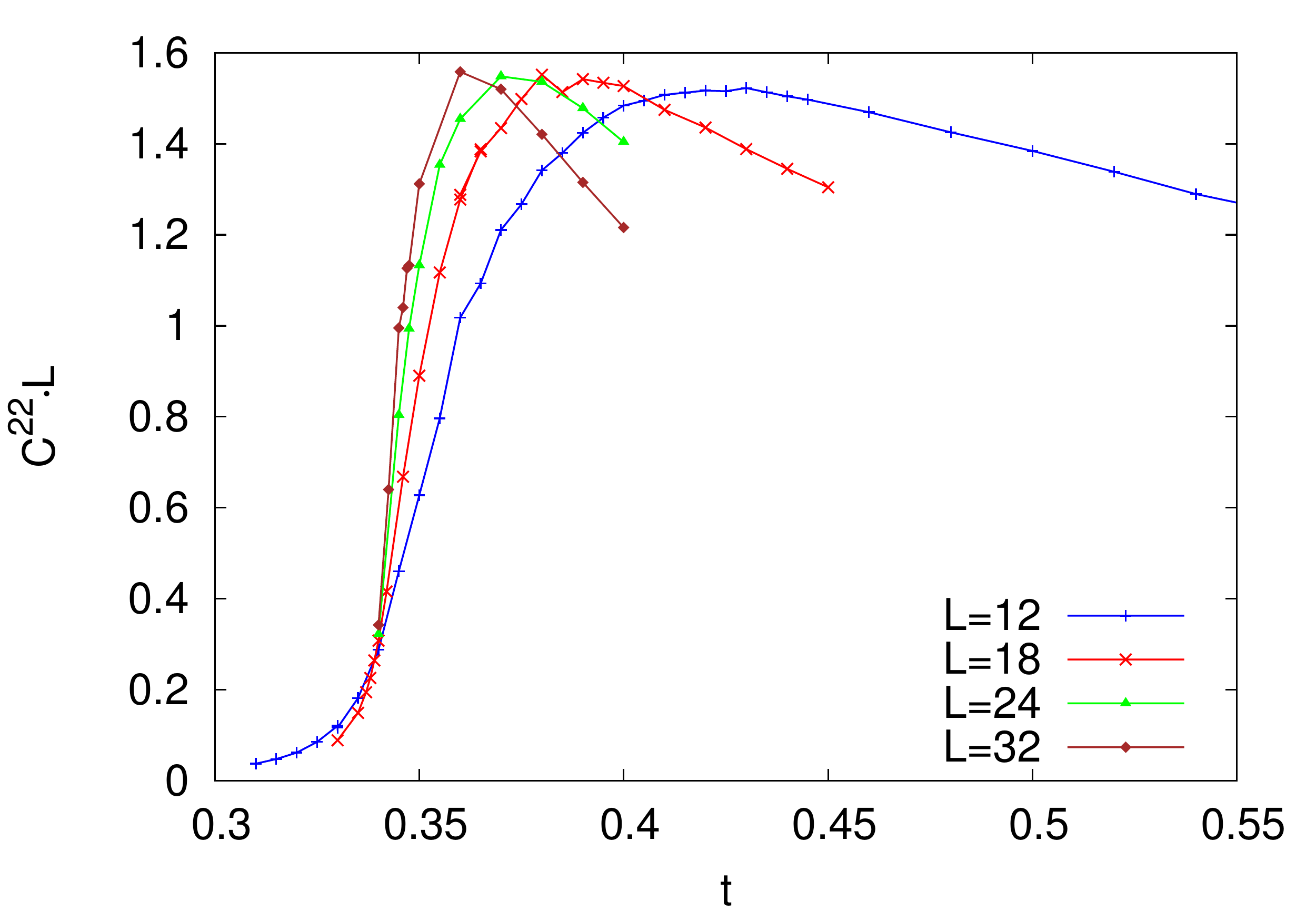}
	\vspace{-0.1in}
	\caption{\label{fig:P} Plot of $C^{22} \cdot L$ along the symmetric line over a broad range of parameters covering phases (0) and (IV).  We expect that $C^{22} \cdot L$ vanishes as $1/L$ for sufficiently large $L$ in both phases but is non-zero at the transition.  Top panel in Fig.~\ref{fig:3} shows selection of the same measurements near the critical point, while Figs.~\ref{fig:5}~and~\ref{fig:6} show our analysis of peak derivatives of this quantity used to extract the critical exponents.
}
\end{figure}


Therefore, we ruled out scenarios (b) and (c). The only possibility left is scenario (a), so we conclude that this critical region contains a multi-critical point where four phases meet.  In the renormalization group language, this multicritical point corresponds to a fixed point with two relevant directions, which in the present case are constrained to lie in the symmetric and anti-symmetric directions as depicted in Fig.~\ref{fig:scenarios}(a).

If we assume that the scenario (a) is correct, we can then deduce the behavior of $C^{22}\cdot L$ at the multicritical point. Recall that $C^{22}\cdot L\sim 1/L$ in phases (0) and (IV), and therefore one might think that it would have this behavior at the critical point as well. This is incorrect, which can be seen by considering the transition along the anti-symmetric direction, from phase (I) to phase (II). In phase (I) $C^{22}\cdot L\sim 1/L$, but in phase (II) it is $\sim L$, and therefore at the critical point we expect $C^{22}\cdot L$ to be independent of the system size. In Fig.~\ref{fig:P} we show the behavior of $C^{22}\cdot L$ along the symmetric line, and we see that it is indeed $\sim 1/L$ in phases (0) and (IV) but that it saturates to a constant value at the critical point. This allows us to use the finite-size scaling form in Eq.~(\ref{scaling}) to extract the critical exponents. Note that the behavior of $C^{22}\cdot L$ in Fig.~\ref{fig:P} also rules out a first order transition between (I) and (II), as for such a transition $C^{22}\cdot L$ would diverge with increasing system size [since $C^{22}$ is constant in phase (II)], while we clearly observe bounded $C^{22}\cdot L$.

\begin{figure}
	\includegraphics[width = 1.05\columnwidth, angle =0]{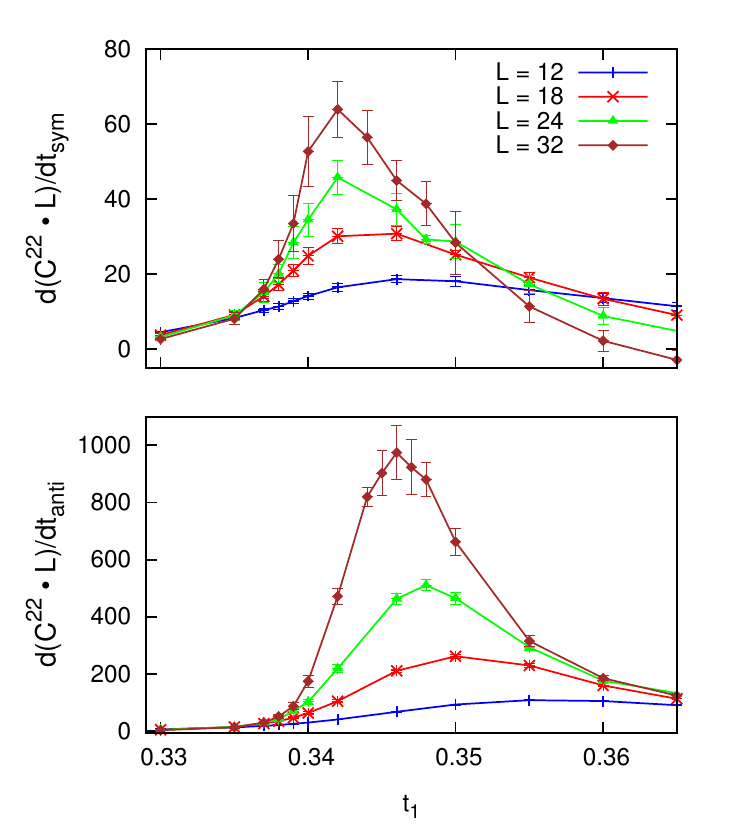}
	\vspace{-0.1in}
	\caption{\label{fig:5} Behavior of the symmetric/anti-symmetric derivative of the current-current correlation $C^{22} \cdot L$, for $\theta = 2\pi/3$. Similar plots are also obtained at other $\theta$. Error bars are estimated by running 20 simulations with different seeds. Peak positions in these plots are taken to be finite-size critical points, and the corresponding peak values are analyzed in Fig.~\ref{fig:6}. The upper panel is essentially the derivative of the curves in Fig.~\ref{fig:P} (and top panel in Fig.~\ref{fig:3}), which themselves were taken along the symmetric line.  However, both the symmetric and anti-symmetric derivatives are evaluated using a more accurate method in Eq.~(\ref{eq:deriv}).
}
\end{figure}

\subsection{Critical Exponents}
We are now in a position to extract critical exponents for the phase transition between the fractional quantum Hall insulator and the trivial insulator. There are two relevant correlation length exponents, $\nu_s$ and $\nu_a$, corresponding to the symmetric and antisymmetric cuts in the phase diagram. We are interested in the behavior of these critical exponents as a function of the parameter $\theta$ in Eq.~(\ref{eq:model}).  
The model with $\theta = 0$ corresponds to two decoupled integer-valued conserved currents. This case can be simply reformulated into the 3D XY model for each species, and it is well-known that the critical exponent is $\nu \approx 0.670$  for this case~\cite{Sorensen}; the phase diagram is simply divided into four regions by two lines, vertical line at $t_1 \approx 0.33325$ and horizontal line at $t_2 \approx 0.33325$.~\cite{short_range3}

We then consider the $\theta$ term in Eq.~(\ref{eq:model}) as a modification to the decoupled 3D XY models, which has qualitative effect as it modifies the critical indices; however, these changes are small for small $\theta$.  For $\theta = 0$, we have $\nu_s = \nu_a \approx 0.67$, and as $\theta$ increases from zero we expect the exponents will deviate from this value.  This also changes how the phase boundaries approach each other as we can see in Fig.~\ref{fig:phase}, since the shapes of the phase boundaries near such a multicritical point are determined by the critical exponents. Renormalization group arguments tell that the phase boundary line has the form $\delta t^{\rm bd}_a \sim (\delta t^{\rm bd}_s)^{\nu_s/\nu_a}$, where $\delta t^{\rm bd}_s$ and $\delta t^{\rm bd}_a$ are deviations from the multicritical point in the symmetric and antisymmetric directions respectively. Therefore, scenario (a) of Fig.~\ref{fig:scenarios} corresponds to $\nu_s > \nu_a$ since the boundary is bending away from the symmetric line, which is what we find from Monte Carlo results.

\begin{figure} 
\vspace{-0.15in}

	\includegraphics[width = 1.01\columnwidth, angle = 0]{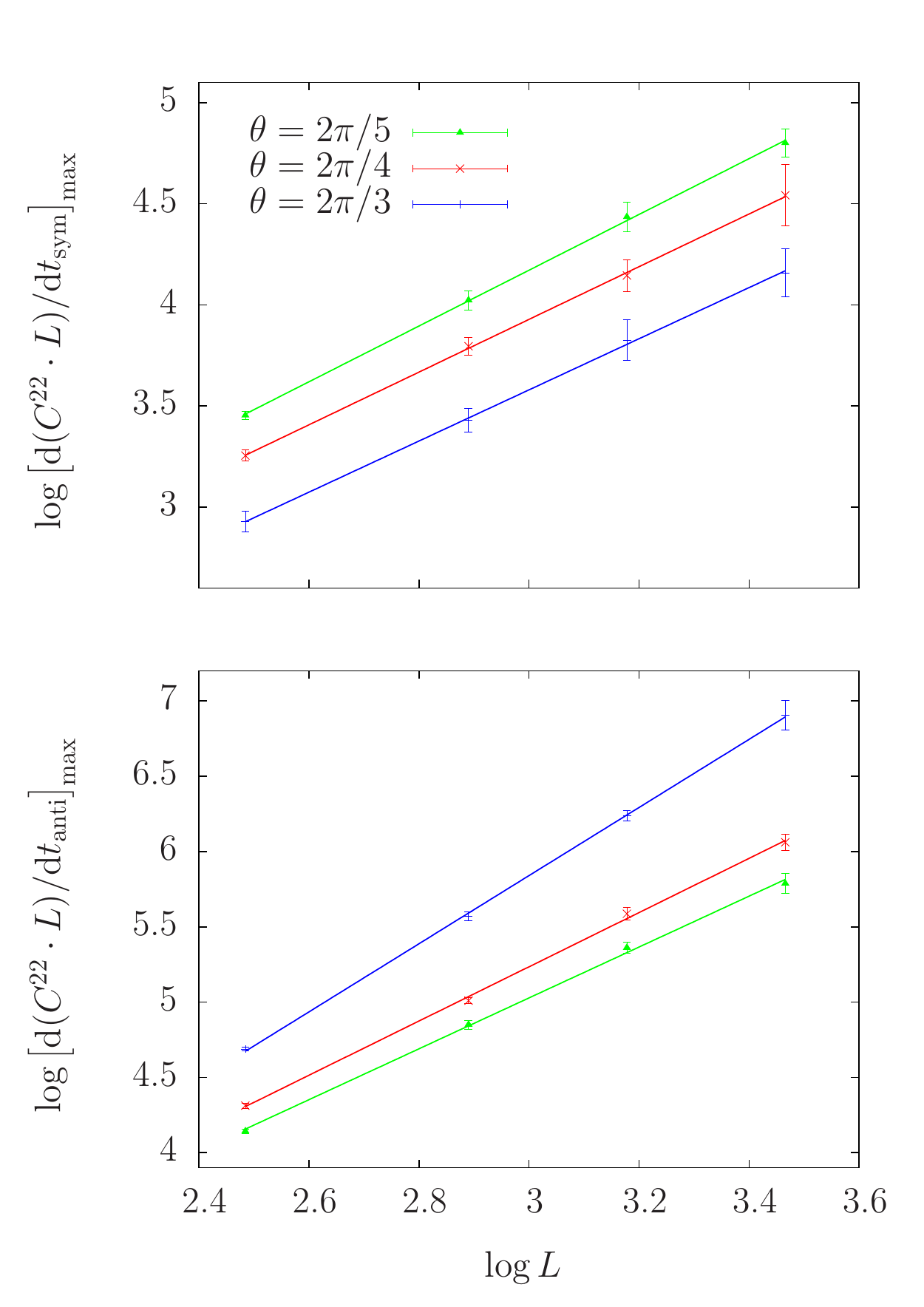}
	\caption{\label{fig:6} Log-log plot of the peaks of the symmetric/anti-symmetric derivatives of the current-current correlation \mbox{$C^{22}\cdot L$} versus system size, extracted from the top/bottom panels in Fig.~\ref{fig:5} for $\theta = 2\pi/3$ and similar analysis for $\theta = 2\pi/4$ and $2\pi/5$. Because $[d(C^{22} \cdot L)/dt]_{\text{max}} \propto L^{1/\nu}$, the slopes of these lines give the inverses of the corresponding critical exponents $\nu$. 
}
\end{figure}

We find that models with $\theta = 2\pi/n$ with integer $n$ always have a phase diagram qualitatively similar to Fig.~\ref{fig:phase}, although the locations of the phase transitions change with $n$.\cite{short_range3} We studied the region in the blue circle for $n = 3, 4, 5$ and concluded that all these cases have multi-critical points as in scenario (a) of Fig.~\ref{fig:scenarios}. We extracted critical exponents by examining the derivatives of the correlation $C^{22}\cdot L$ with respect to symmetric/antisymmetric deviations, as shown in Fig.~\ref{fig:5} for the case of $\theta=2\pi/3$. We expect the peak values of this quantity to be proportional to $L^{1/\nu}$, with appropriate $\nu = \nu_s$ or $\nu_a$, as discussed after Eq.~(\ref{scaling}). Therefore we plot these peaks as a function of system size (on a log-log plot) in Fig.~\ref{fig:6}. The slope of the resulting lines is equal to $1/\nu_s$ or $1/\nu_a$.

\begin{figure} 

	\includegraphics[height = 1.06\columnwidth, angle = 0]{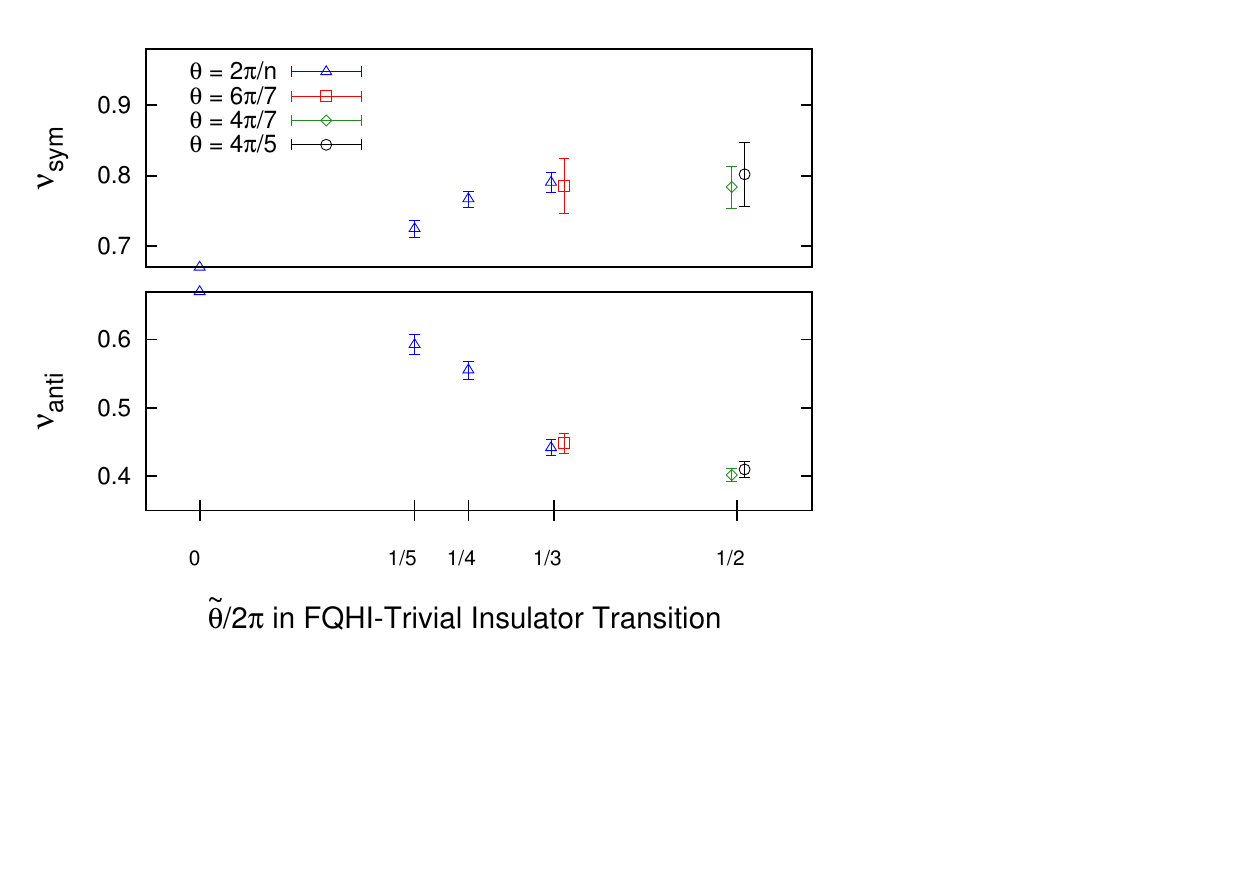}
	\vspace{-1in}
	\caption{\label{fig:7} Critical exponents for phase transitions from FQHI phases to the trivial phase, where $\tilde{\theta}$ represents the effective mutual statistics of quasiparticles condensing in such phase transitions. Our models with $\theta = 2\pi/n$ have transitions from FQHI phases with $\sigma^{12}_{xy} = 2\frac{1}{n}$ to the trivial phase with $\sigma^{12}_{xy} = 0$; as remarked, quasiparticles with mutual statistics $\tilde{\theta} = \theta$ are condensing in these cases.
	Our models with $\theta = 4\pi/7$ and $4\pi/5$ allow us to study transitions with $\tilde{\theta} = \pi$.  Specifically, we study the transition from the phase with $\sigma^{12}_{xy} = 2\frac{1}{2}$ to the trivial insulator realized in these cases, cf.~Fig.~\ref{fig:phased4pi5}; at this transition, quasiparticles with mutual statistics $\tilde{\theta} = \pi$ are condensing.  We find that the nature of this transition appears to be different from that in the model with bare $\theta = \pi$, where a first-order segment is found along the symmetric line, while in the present realization we find a direct transition.
	Finally, our model with $\theta = 6\pi/7$ gives us another instance of the transition with $\tilde{\theta} = 2\pi/3$, which is the transition from the FQHI phase with $\sigma^{12}_{xy} = 2\frac{1}{3}$ to the trivial phase realized in this case; we find that the critical exponents are consistent with those in the case with bare $\theta = 2\pi/3$.
	}
\end{figure}

Figure~\ref{fig:7} shows the critical exponents $\nu_s$ and $\nu_a$ for each value of $\theta$ extracted from the analysis in Fig.~\ref{fig:6}. At $\theta = 0$, both critical exponents have the same value 0.670, which is the critical exponent of the 3D XY model. 
We can see that as $\theta = 2\pi/n$ increases, our symmetric exponents increase while the antisymmetric exponents decrease.  The exponents clearly deviate from the decoupled case and vary smoothly, suggesting novel criticality with continuously varying exponents, presumably due to strictly marginal statistical interactions of the condensing particles.
[In an earlier paper\cite{Loopy} we considered the same model with $\theta = \pi$ and found instead scenario (b) in Fig.~\ref{fig:scenarios}  with first-order behavior on the segment along the symmetric line.  While the multi-critical point scenario in the present work is qualitatively different, we note that such a first-order behavior can manifest itself as if $\nu_a = 1/3$, and our antisymmetric exponents for increasing $\theta$ may be moving towards such value.]

\section{Phase Diagrams with different $\theta$}
\label{sec:general}
So far in this work we have considered the model with $\theta = 2\pi/n$, which has one FQH insulator phase and a transition to trivial insulator.
As described in the Appendix, we can also 
estimate the structure of the phase diagram in our model at general $\theta$, which we can then check and study in detail using Monte Carlo simulations.

\begin{figure}
	\hspace{-0.1in}\includegraphics[width = 1.15\columnwidth]{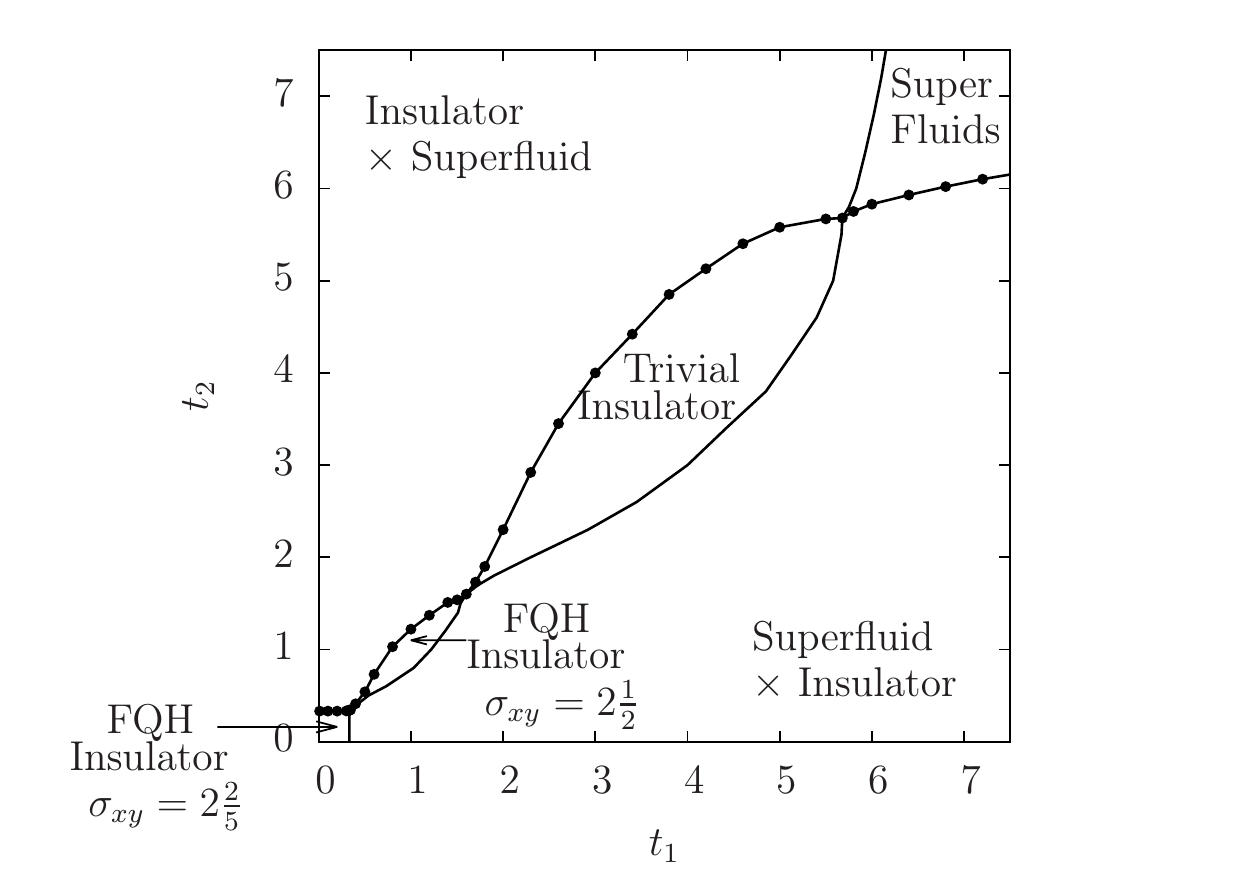}
        \caption{\label{fig:phased4pi5} The phase diagram for the model with $\theta = 4\pi/5$.  Unlike $\theta = 2\pi/n$ cases, it has two distinct fractionalized phases, with physical $\sigma^{12}_{xy} = 2\frac{2}{5}$ and $\sigma^{12}_{xy} = 2\frac{1}{2}$.  Points show where the locations of phase transitions were determined by finding peaks in the specific heat.
Qualitatively similar phase diagrams are obtained for family of $\theta = 2\pi \frac{n_1}{n_1 n_2 + 1}$, see text for details.
}
\end{figure}

As an example, from such studies we determine that the model with $\theta = 2\pi \frac{n_1}{n_1 n_2 + 1}$ has two distinct FQHI phases:  The first phase is in the lower left corner of the phase diagram in the $t_1$-$t_2$ parameter space, cf.~Fig.~\ref{fig:phased4pi5}; this phase has physical quantum Hall conductivity $\sigma^{12}_{xy} = 2\frac{n_2}{n_1 n_2 + 1}$ and the gapped quasiparticles are simply the $J$'s of our starting model Eq.~(\ref{eq:model}).  The second FQHI phase is adjacent to the first along the diagonal and occurs when the $J$'s condense; this phase has physical $\sigma^{12}_{xy} = 2\frac{1}{n_1}$ and its quasiparticles are vortices in $J$'s and have long-distance mutual statistics $\theta_{\rm dual} = (2\pi)^2/\theta = 2\pi/n_1$ (modulo $2\pi$).  Note that the $n_1$ are $n_2$ are not restricted to be even or odd, and so the rational numbers accessible to these bosonic models are different from those obtained in the hierarchy picture of the fractional quantum Hall effect of bosons in strong magnetic field.  Naturally, when $\theta$ is a more complicated rational number, a phase diagram with more fractionalized phases can arise.

Let us consider numerical example with $\theta = 4\pi/5$ which corresponds to $n_1 = n_2 = 2$.  We find the phase diagram in Fig.~\ref{fig:phased4pi5} with the FQHI phases as described above.  We can extract the critical exponents of the phase transitions in this diagram.  In the lower corner we have a transition from $\sigma^{12}_{xy} = 2 \frac{2}{5}$ to $\sigma^{12}_{xy} = 2 \frac{1}{2}$.  We find that this is a continuous phase transition between two FQHI phases, for which we estimate $\nu_s = 0.78 \pm 0.03$ and $\nu_a = 0.37 \pm 0.01$. In the middle, we have a transition from the FQHI phase with $\sigma^{12}_{xy} = 2 \frac{1}{2}$ to a trivial insulator.  The gapped quasiparticles in this FQHI phase and which are condensing at this transition are not the original quasipartices, they are some new quasiparticles with effective mutual statistics $\tilde\theta=\pi$.  The critical exponents for this transition have been plotted in Fig.~\ref{fig:7}, alongside with the other FQHI-to-trivial transition critical exponents.

We also studied the cases of $\theta = 4\pi/7$ (corresponding to $n_1 = 2, n_2 = 3$) and $6\pi/7$ (corresponding to $n_1 = 3, n_2 = 2$), which have FQHI phases with $\hall = 2\frac{3}{7}, 2\frac{1}{2}$ and $\hall = 2\frac{2}{7}, 2\frac{1}{3}$ respectively.  We determined the critical exponents for transitions from the second (upper) FQHI phase to the trivial insulator, and plotted the results in Fig.~\ref{fig:7}.  These are realizations of  phase transitions corresponding to effective $\tilde\theta = \pi$ and $2\pi/3$ respectively, which we have already encountered in the $\theta = 4\pi/5$ and $2\pi/3$ models.  Though the different models are realizing transitions which are presumably in the same universality class determined by $\tilde{\theta}$ for the condensing mutual anyons, they have different microscopic interactions between these quasiparticles (see Appendix).  This provides a useful check on the universality and our procedure, as we can see from Fig.~\ref{fig:7} that the transitions in the different models but realizing the same $\tilde{\theta}$ have the same critical exponents.

\section{Discussion}
\label{sec:concl}
In this work, we studied phase transitions between a bosonic ``fractional quantum Hall insulator'' phase and a ``trivial insulator,'' and between different bosonic FQHI phases. We found that the phase transition takes place at a second-order multi-critical point, and we extracted critical exponents using finite-size scaling.

The physical problems studied in this work are conceptually related to those considered for more familiar fractional quantum Hall systems in strong magnetic fields---namely phase transitions involving topological phases.\cite{WenWu1993, ChenFisherWu1993}  Indeed, let us first write the continuum field theory that corresponds to our lattice problems:
\begin{widetext}
\begin{eqnarray}
\label{CSQft}
S_{\rm eff}[\Psi_1, \Psi_2, \vec{\alpha}_1, \vec{\alpha}_2] &=&
\int d^3r \left[g_1(\vec{\nabla} \times \vec{\alpha}_1)^2 + g_2(\vec{\nabla} \times \vec{\alpha}_2)^2
-\frac{i}{\tilde{\theta}} \vec{\alpha}_1 \cdot (\vec{\nabla} \times \vec{\alpha}_2) \right] \\
&+& \int d^3r \left[\gamma_1|(\vec{\nabla} - i\vec{\alpha}_1)\Psi_1|^2 + \gamma_2|(\vec{\nabla} - i\vec{\alpha}_2)\Psi_2|^2 + m_1|\Psi_1|^2 + m_2|\Psi_2|^2 + ({\rm quartic~terms})\right] ~.\nonumber
\end{eqnarray}
\end{widetext}
Here $\Psi_1$ and $\Psi_2$ are complex-valued matter fields in continuum corresponding to integer-valued current variables on the lattice. These matter fields are coupled to gauge fields $\vec{\alpha}_1$ and $\vec{\alpha}_2$ respectively, and these gauge fields have a mutual Chern-Simons term characterized by a statistical angle $\tilde{\theta}$, encoding precisely $\tilde{\theta}$ statistical interaction between the two particle species.  $g_{1/2}$, $\gamma_{1/2}$, and $m_{1/2}$ represent some effective parameters .  We are interested in the transition along the symmetric line ($g_1 = g_2$, $\gamma_1 = \gamma_2$, $m_1 = m_2$, etc.), where both matter fields $\Psi_1$ and $\Psi_2$ are condensing simulatenously.

When $\Psi$'s represent the $J$ currents in Eq.~(\ref{eq:model}), and $\tilde\theta = \theta$ is the mutual statistics angle for those currents, then the field theory in Eq.~(\ref{CSQft}) describes the transition out of the bottom left corner phase in phase diagrams such as Figs.~\ref{fig:phase} and \ref{fig:phased4pi5}. More generally, $\Psi$'s can represent gapped quasiparticles in any other bosonic FQHI phase. In that case $\tilde\theta$ will represent the mutual statistics of those quasiparticles. For example, in the phase diagram in Fig.~\ref{fig:phased4pi5}, which is representative of our model Eq.~(\ref{eq:model}) with $\theta = 2\pi \frac{n_1}{n_1 n_2 + 1}$, the transition from the $\sigma^{12}_{xy} = 2\frac{1}{n_1}$ phase to the trivial insulator can be described by Eq.~(\ref{CSQft}) with $\tilde{\theta} = 2\pi \frac{1}{n_1}$, where $\Psi$'s represent the gapped quasiparticles in this phase (see the Appendix for more details).

A version of Eq.~(\ref{CSQft}) with only one species of bosons with self-statistics has been used to describe a transition between a fractional quantum Hall state and a trivial Mott insulator, as well as transitions between different fractional quantum Hall plateaus.\cite{WenWu1993, ChenFisherWu1993} It was found that the Chern-Simons term is a marginal perturbation and therefore critical exponents should depend on $\tilde{\theta}$.  However, finding these critical exponents analytically has been possible only in some artificial limits with large number of fields, while there are no controlled results for the experimentally relevant number of fields. The critical exponents for our model can be computed using unbiased numerics, and it would be interesting to compare them with results of applying the analytical methods of Refs.~\onlinecite{WenWu1993, ChenFisherWu1993} to our two-species case.

Our study was possible due to the existence of a lattice model of the fractionalized phases which can be studied in sign-free Monte Carlo simulations. This lattice model could also be used to study other properties of these exotic phases, such as their response to disorder or their entanglement properties.  


\acknowledgments

This research was supported by the National Science Foundation through grant DMR-1206096, and by the Caltech Institute of
Quantum Information and Matter, an NSF Physics Frontiers Center with the support of the Gordon and Betty Moore foundation. SG was supported by an NSERC PGS fellowship and JL was supported by an NSF SURF fellowship.


\appendix

\section{Details of Variable Transformations}
In Sec.~\ref{sec:general} we studied values of $\theta$ which did not have the form $2\pi/n$, and found multiple fractionalized phases.  This motivates the question of what phases can arise for a given value of $\theta$.  The definitive way to answer this is to perform Monte Carlo simulations for the value of $\theta$ in question.  However, as simulations can be time-consuming, we have also developed a heuristic for determining which phases are likely to occur, which is outlined in this Appendix.  We have tested this heuristic numerically for several values of $\theta$ and found that it gives the correct sequence of phases and also reasonable estimates of transition points.

To understand the phase diagrams with different $\theta$, we will exploit ``modular transformation'' approach developed in Ref.~\onlinecite{Gen2Loops}.  The key idea is to rewrite the partition sum coming from the action in Eq.~(\ref{eq:model}) in terms of new variables that also represent some bosons with mutual statistics, but the interactions between them will be different from the original problem.  This technique was useful in the main text because it allowed us to interpret our results in terms of the ``elementary bosons'' of the problem, which we will define explicitly below, instead of the particles in Eq.~(\ref{eq:model}).  It is also useful because, for given model parameters, we can find new variables which see large repulsive interactions and are therefore gapped. These variables are then gapped quasiparticles of a phase occuring for the specified model parameters, and it is then easy to analyze the properties of the phase in terms of these gapped quasiparticles, since their correlation functions vanish exponentially at long distances.

We now demonstrate how to reformulate Eq.~(\ref{eq:model}) in terms of new variables.  We first express the action in momentum ($k$)-space as follows:
\begin{eqnarray}
S &=& \frac{1}{2} \sum_k \left[ v_1(k) |\vec{J}_1(k)|^2 + v_2(k) |\vec{J}_2(k)|^2 \right] \nonumber \\
&+& i \sum_k \theta(k) \vec{J}_1(-k) \cdot \vec{p}_2(k) ~. \label{action}
\end{eqnarray}   
Starting from this action, we perform a ``duality'' transformation $\vec{J}_1 \to \vec{Q}_1$ so that the partition sum is expressed in terms of new variables $(\vec{Q}_1, \vec{J}_2)$.  The $\vec{Q}_1$ variables are vortices of the original bosons $\vec{J}_1$, and they are integer-valued conserved currents defined on the links of a cubic lattice which is dual to the lattice the $\vec{J}_1$ live on.  A derivation of this duality for our models is given in the Appendix of Ref.~\onlinecite{short_range3}, and the approach is physically equivalent to the $(2+1)$-dimensional boson-vortex duality known in the literature.\cite{PolyakovBook, Peskin1978, Dasgupta1981, FisherLee1989, LeeFisher1989, artphoton}

Once the duality transformation has been performed, we have an action in terms of variables $\vec{Q}_1$ and $\vec{J}_2$, both of which are integer-valued conserved currents living on the links of the same lattice.  We can therefore perform the following $SL(2, \mathbb{Z})$ (modular) transformation:
\begin{eqnarray}
\vec{F}_1 &=& a\vec{Q}_1 - b\vec{J}_2 \\  
\vec{G}_2 &=& c\vec{Q}_1 - d\vec{J}_2 ~, 
\end{eqnarray}   
where $a, b, c, d \in \mathbb{Z}$ and $ad - bc = 1$.
Under this transformation, the new variables $\vec{F}_1$ and $\vec{G}_2$ are independent integer-valued currents.  The above condition requires that $c$ and $d$ are mutually prime, which will be assumed throughout.  Finally, we perform another duality transformation, $\vec{F}_1 \to \vec{G}_1$, to express the action in terms of $(\vec{G}_1, \vec{G}_2)$.  We have now completed the modular transformation, obtaining the following action:
\begin{eqnarray}
S[\vec{G}_1, \vec{G}_2] &=& \frac{1}{2} \sum_k [ v_{G1}(k) |\vec{G}_1(k)|^2 + v_{G2}(k) |\vec{G}_2(k)|^2 ] \nonumber \\
&+& i \sum_k \theta_G(k) \vec{G}_1(-k) \cdot \vec{p}_{G2}(k) ~.
\end{eqnarray}
The new loops have intra-species interactions with potentials
\begin{equation}
v_{G1/2}(k) = \frac{ (2\pi)^2 v_{1/2}(k) }{ [2\pi d + \theta(k) c ]^2 + v_1(k) v_2(k) |\vec{f_k}|^2 c^2 } ~, \label{poten}
\end{equation}
and also an effective inter-species statistical interaction described by
\begin{equation}
\frac{ \theta_{G}(k) } {2\pi} = \frac{ [2\pi b + \theta(k) a][2\pi d + \theta(k) c] + v_1(k)v_2(k) |\vec{f_k}|^2 ca }{ [2\pi d + \theta(k) c ]^2 + v_1(k) v_2(k) |\vec{f_k}|^2 c^2 } ~. \label{theta}
\end{equation}
In the above equations, we defined $|\vec{f}_k|^2 = \sum_\mu (2 - 2\cos k_\mu)$, which vanishes as $|\vec{f}_k|^2 \approx |\vec{k}|^2$ for small $k$.  We note that $\theta_{G}(k)$ depends on $k$ even when the original $\theta(k)$ is momentum-independent.

\begin{figure}
	\includegraphics[height=0.9\columnwidth, angle=270]{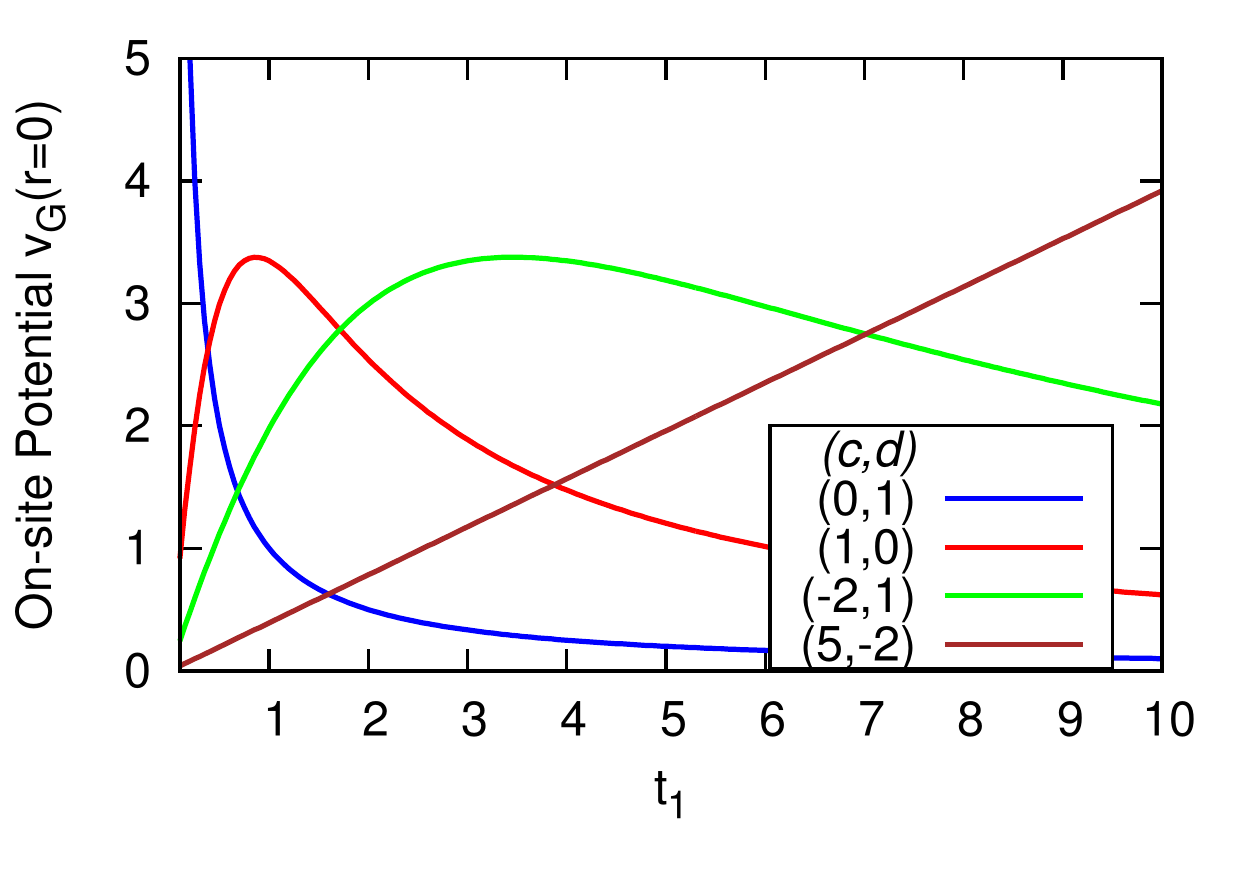}
	\caption{\label{fig:Vonsite4pi5} Strength of the on-site potential for quasiparticles obtained by a modular transformation labelled $(c, d)$.  The analysis is for $\theta = 4\pi/5$ (compare to Fig.~\ref{fig:phased4pi5}) and along the symmetric line $t_1 = t_2 = t$; the on-site potential is obtained by Fourier transforming Eq.~(\ref{poten}) to real space. 
For given $t$, we find $(c, d)$ that give the largest on-site potential and interpret the corresponding quasiparticles as gapped.
The intersection points between potentials where the optimal $(c, d)$ changes gives rough estimates for the phase transitions in Fig.~\ref{fig:phased4pi5}.
}
\end{figure}

Note that $v_{G}(k)$ explicitly does not depend on $a$ or $b$, while $\theta_G(k)$ appears to.  However, because of the condition $ad - bc = 1$, different such choices $(a, b)$ and $(a', b')$ are related by $a' = a + cm$, $b' = b + dm$, with $m$ an integer.  This changes $\theta_G(k)$ by $2\pi m$, and we can see from Eq.~(\ref{eq:model}) that this does not affect the statistical mechanics of the $\vec{G}_{1/2}$ particles. From now on we will therefore label our modular transformations by $(c,d)$ only. 

We can perform the above transformation for arbitrary $(c, d)$, but most such transformations lead to new variables which do not help us to understand the physics of our model. In this work the useful transformations are the ones in which the new variables are gapped, and the ones in which the new variables can be thought of as ``elementary bosons.''

The ``elementary bosons'' are integer-valued conserved currents that, unlike the currents in Eq.~(\ref{eq:model}), experience only local interactions. The action in terms of these bosons can be derived from the path integral of a local Hamiltonian.\cite{FQHE}
For a model with $\theta/(2\pi) = p/q$ with mutually prime $p$ and $q$, we can find the elementary bosons by choosing, e.g., $a_{\rm elem} = q, b_{\rm elem} = -p$, and $c_{\rm elem}, d_{\rm elem}$ to satisfy $a_{\rm elem} d_{\rm elem} - b_{\rm elem} c_{\rm elem} = 1$.  
This still leaves some arbitrariness in the choice of $(c_{\rm elem}, d_{\rm elem})$, which we fix by requiring $c_{\rm elem}$ to be as small as possible.  
It is easy to see from Eq.~(\ref{theta}) that the resulting variables  $\vec{G}_{1, {\rm elem}}, \vec{G}_{2, {\rm elem}}$ have $\theta_G(k) \sim |k|^2$ at long wavelengths, so the interaction between these variables encoded in the $\theta_G$ is actually short-ranged (note that the interactions encoded in the $v_{G,1/2}$ are also short ranged).  It is the local nature of these statistical interactions which motivates us to call these variables the elementary bosons.

We can determine the phase diagram for our models semi-quantitatively by determining which loop variables are gapped.  For a given value of $\theta$, we can determine the strength of the loop-loop interactions as a function of $t = t_1 = t_2$, focusing on the symmetric line for concreteness.  We can do this for different values of $c$ and $d$, and assume that the gapped quasiparticles at a given value of $t$ are the ones which see the largest repulsive interactions, which we can quantify roughly by looking at the on-site potential $v_G(r = 0)$ in real space.

Once we have found the values of $(c, d)$ which give the largest on-site potential, we want to determine the properties of the phase when this kind of quasiparticle is gapped. 
We can read off the statistical interactions of these quasiparticles directly from the long-distance behavior of $\theta_G(k)/(2\pi) \approx  [2\pi b + \theta a]/[2\pi d + \theta c]$.  Furthermore, we can determine the properties of the phase in terms of the elementary bosons defined earlier.
Indeed, we can determine current-current correlations of elementary bosons in any phase, once we know what $(c, d)$ give gapped quasiparticles in that phase.\cite{Gen2Loops}

Applying such an analysis in the general case, we find that for the phase in the upper right corner occurring for very small $v_{1,2}$, its gapped quasiparticles are obtained by using $c = -q, d = p$, again assuming rational $\theta/(2\pi) = p/q$ with mutually prime $p$ and $q$.  
We find that superfluid stiffnesses of the elementary bosons are large in this phase, therefore it is a superfluid.
The gapped quasiparticles in this case have long-range interactions [$v_G(k) \sim 1/k^2$ corresponding to $v_G(r) \sim 1/r$ in real-space] and are interpreted as vortices in the superfluid.
The phase diagonally adjacent to the superfluid phase has vanishing superfluid stiffness and vanishing Hall conductivity, and this is why we claim that it is a trivial insulator; the gapped quasiparticles here are precisely the elementary bosons and have only short-range interactions (no statistical interactions on long distances).
For the other phases, we find a quantized Hall conductivity, and gapped quasiparticles with short-range intra-species interactions and non-zero inter-species statistical interactions on long distances.

We now give an example of this technique for the case where $\theta/(2\pi) = n_1/(n_1 n_2 + 1)$, with $n_1$ and $n_2$ arbitrary integers.  The elementary bosons in this case are obtained using $a_{\rm elem} = n_1 n_2 + 1$, $b_{\rm elem} = -n_1$, $c_{\rm elem} = -n_2$, $d_{\rm elem} = 1$.
Figure~\ref{fig:Vonsite4pi5} shows the onsite potential as a function of $t$ for a variety of different $(c, d)$ for $n_1 = n_2 = 2$, which corresponds to the phase diagram in Fig.~\ref{fig:phased4pi5}.  We see that at small $t$, the original $\vec{J}_{1/2}$ variables are gapped: $(a, b, c, d) = (1, 0, 0, 1)$ gives $\vec{G}_{1/2} = -\vec{J}_{1/2}$; translated to physical variables, this phase has $\sigma^{12}_{xy} = 2\frac{n_2}{n_1 n_2 + 1}$. 
As we increase $t$, we expect to condense the $\vec{J}_{1/2}$ variables, and so we consider the vortices of these variables which can be obtained by applying our transformation with $(a, b, c, d) = (0, -1, 1, 0)$; we can extract the physical Hall conductivity $\sigma^{12}_{xy} = 2\frac{1}{n_1}$ in this case.
At even larger $t$ the gapped quasiparticles are the elementary bosons, $(c, d) = (c_{\rm elem}, d_{\rm elem})$, so this phase is the trivial insulator. 
Finally, at very large $t$ we have $c = -(n_1 n_2 + 1), d = n_1$.  As discussed earlier, this phase is the superfluid of the elementary bosons.

Note that since we have considered only the on-site components of $v_G$, and also neglected $\theta_G$, the exact locations of the phase transitions will be different from those that one would infer from Fig.~\ref{fig:Vonsite4pi5}. Despite this, we have tested this method for $\theta = 4\pi/5$ (see Fig.~\ref{fig:phased4pi5}), $\theta = 4\pi/7$, and $\theta = 6\pi/7$, and found that it correctly predicts which phases exists and even approximately predicts the transition points. This is not too suprising as in real space the potentials decay rapidly [numerically we find $v_G(r=1)/v_G(r=0)<0.1$ for the considered cases], and the onsite potential alone is large enough to gap a single species of bosons.\cite{Sorensen}

\bibliographystyle{aipnum4-1}
\bibliography{thesis.bib}

\end{document}